\documentclass{PoS}

\title{ PHENIX results on system size dependence of J/$\psi$ nuclear modification in $p$,~$d$, $^{3}$He+A collisions at $\sqrt{s_{NN}}$=200 GeV}

\ShortTitle{J/$\psi$ production in small systems}

\author{\speaker{J. Matthew Durham, for the PHENIX Collaboration }\\
        Los Alamos National Laboratory, Los Alamos, NM 87545 USA\\
        E-mail: \email{durham@lanl.gov}}

\abstract{The dissociation of quarkonia in the medium created in heavy ion collisions is still one of the most debated topics in the heavy ion community. Progress in understanding the effect relies on a broad range of measurements in collisions with different nuclear sizes, covering a broad rapidity range. This presentation will report the $J/\psi$ measurements performed by the PHENIX collaboration at forward and backward rapidities of 1.2 $< |y| <$ 2.2 in $p$+Al, $p$+Au, and $^3$He+Au collisions at $\sqrt{s_{NN}}$ = 200 GeV, and the corresponding nuclear modification factors. }

\FullConference{International Conference on Hard and Electromagnetic Probes of High-Energy Nuclear Collisions\\
		30 September - 5 October 2018\\
		Aix-Les-Bains, Savoie, France}

\begin{document}

\section{Introduction}

	Measurements of the bound states of charm quarks produced in collisions involving nuclei can provide information on a wide range of effects in both the early and late stages of the event.  At the center of mass energies probed at the Relativistic Heavy ion Collider (RHIC), the production of primordial $c\bar{c}$ pairs usually proceeds through gluon fusion, making charm quark production sensitive to modifications of the gluon distribution in the nucleus \cite{RamonaOldShadowing}.  After formation the pair must propagate through nuclear matter, where the quarks can experience energy loss \cite{CNM_Eloss} or multiple scattering which can affect the observed transverse momentum distributions of heavy flavor hadrons \cite{Cronin}.  Such interactions can also disrupt the heavy quark precursor state before it has hadronized, resulting in quarkonia suppression.  Even after exiting the nucleus and projecting onto a final state, the fully-formed quarkonia meson can be disrupted through interactions with co-moving hadrons \cite{Capella}. In collisions where quark-gluon plasma is formed, quarkonia suppression due to color screening is expected to be a large effect \cite{MatsuiSatz}.
	
The PHENIX collaboration has previously analyzed data on quarkonia production in $d$+Au collisions at $\sqrt{s_{NN}}$ = 200 GeV \cite{PPG109, PPG125} to study these phenomena.  In this small collision system, quarkonia suppression via color screening in a deconfined plasma is not expected to be the dominant effect.  There have since been multiple calculations which can describe these results, by incorporating the previously described effects to varying degrees \cite{RamonaShadowing,MaCGC,Arleo2013,Ferreiro_comovers}.  The existing data are not sufficient to discriminate between these calculations and allow us to quantify the influence of each suppression mechanism.  Therefore, additional measurements needed.

To this end, the PHENIX collaboration has measured $J/\psi$ production in $p$+Al, $p$+Au, and $^3$He+Au collisions at $\sqrt{s_{NN}}$ = 200 GeV.  By changing the target nucleus (Al versus Au), the level of modification to the nucleon's parton distribution function is expected to vary, with larger effects expected in the larger Au nucleus.  The path length through the nucleus that the charm quarks travel also changes with nuclear size, so energy loss may be affected.  This allows us to experimentally vary the initial state effects that charm quarks may be subject to.  By changing the projectile between $p$,~$d$, and $^3$He, we can experimentally vary the number of produced particles that co-move with the charmonium state outside the nucleus \cite{PPG221}, which may affect the level of breakup that occurs in the late stages of the collision.

\section{Measurement}

The PHENIX detector has two muon spectrometer arms \cite{MuonNIM}, which cover the forward and backward rapidity intervals 1.2 $ < |y|< $ 2.2 and can measure $J/\psi$ through decays to $\mu^{+} \mu^{-}$ down to $p_{T}$ = 0.  Muons produced in the collision travel through a hadron absorber and into a set of wire chambers that provide muon tracking.  Further tracking and muon identification are provided by the MuID, which is layers of streamer tubes interleaved with additional steel absorber panels.  To be considered in this analysis, muons must pass through all absorber material and be detected in the fourth layer of MuID tubes.  The acceptance and efficiency of the detectors are found by running a full GEANT4 \cite{GEANT4} simulation of the PHENIX muon system.

The $^3$He+Au data set was recorded in 2014, and the $p$+Al and $p$+Au samples were recorded in 2015, all at the same center of mass energy per nucleon of 200 GeV.  The two muon spectrometer arms allow data to be recorded in both the forward and backward rapidity ranges simultaneously.  PYTHIA simulations show that in the forward (p,d, or$^3$He -going direction), $J/\psi$ production samples an $x$ range in the target nucleus of $x\approx$ 5$\times$10$^{-3}$, which is in the shadowing region of the gluon parton distribution function.  In the backwards direction, the $x$ range sampled is in the anti-shadowing region near $x\approx$ 8$\times$10$^{-2}$.

\section{Results and Discussion}

\begin{figure}[htbp]
    \centering
    \includegraphics[width=0.495\textwidth]{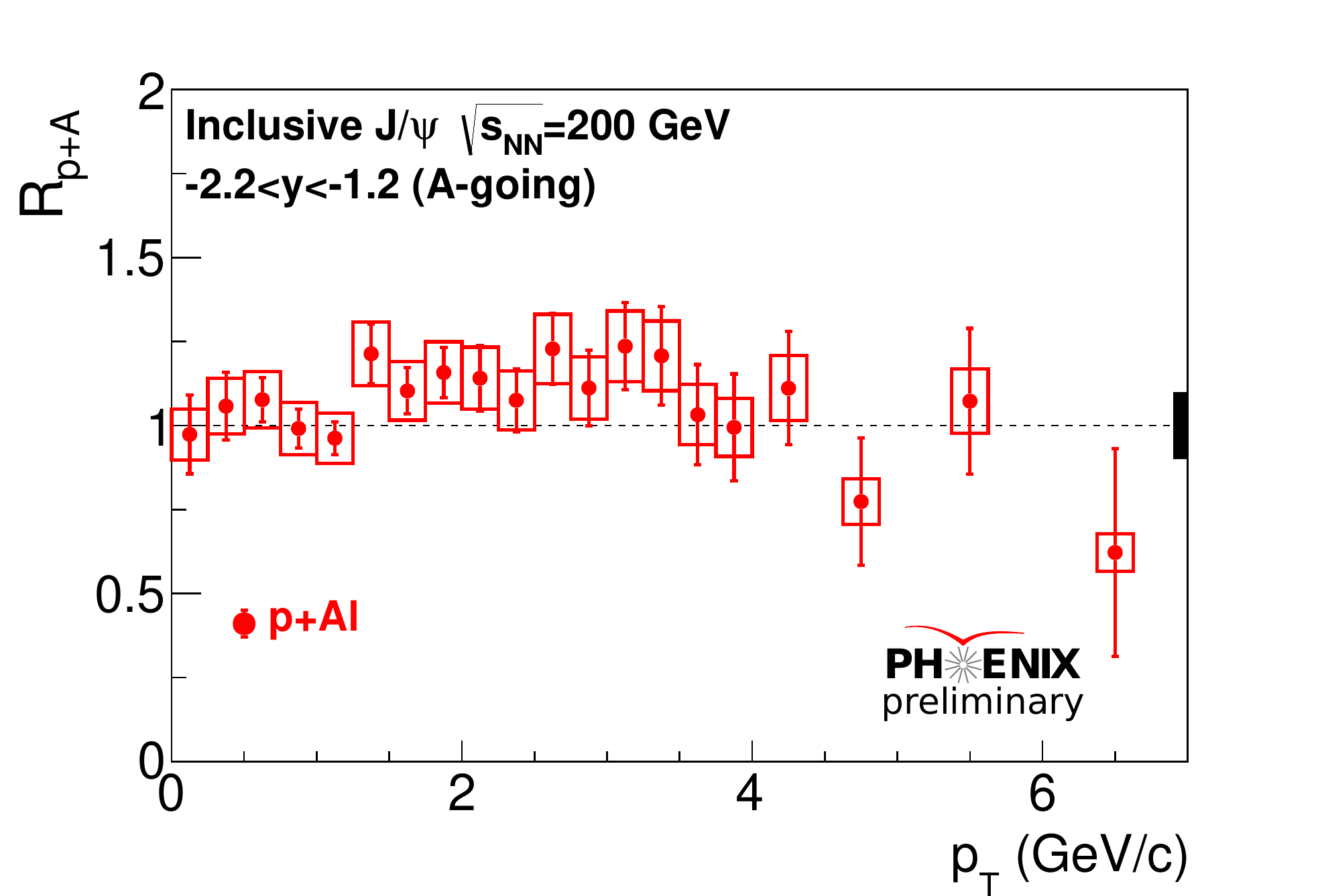}
    \includegraphics[width=0.495\textwidth]{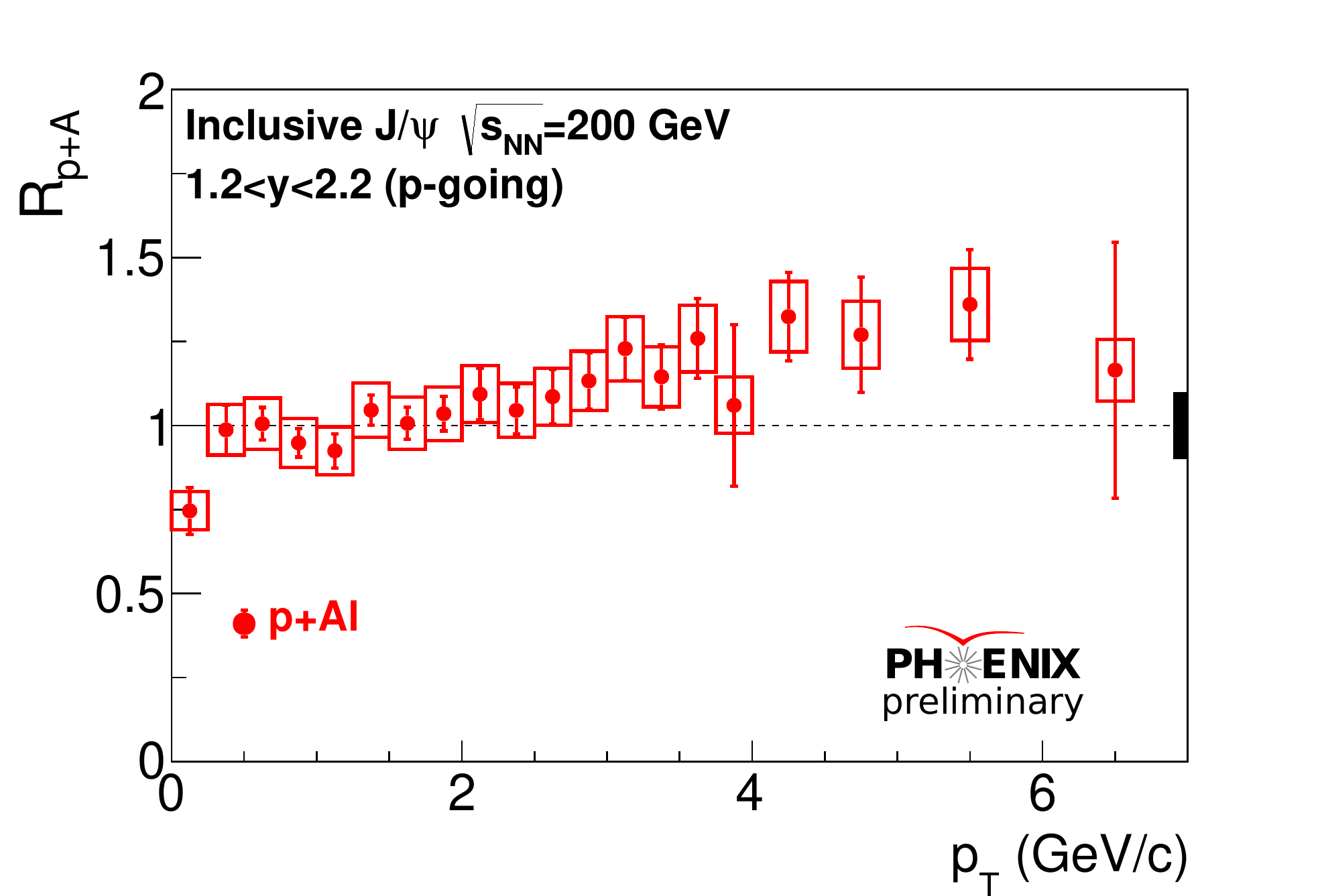}
    \caption{The nuclear modification factor of $J/\psi$ mesons produced at backward (left) and forward (right) rapidity in $p$+Al collisions at 200 GeV.}
    \label{fig:RpAl}
\end{figure}

The $p_{T}$ dependence of the nuclear modification factor of $J/\psi$ produced in minimum bias $p$+Al collisions is shown in Fig. \ref{fig:RpAl}.  The error bars (boxes) represent the statistical (systematic) uncertainty on the points, and the global systematic due to the uncertainty on the $p+p$ denominator is shown as a black box around unity.  The backward and forward rapidity results are shown in the left and right, respectively.  In both cases, we see some evidence for a small amount of enhancement, although within uncertainties the nuclear modification factor is consistent with one.  In this small system, there are no major effects observed on $J/\psi$ production.

\begin{figure}[htbp]
    \centering
    \includegraphics[width=0.495\textwidth]{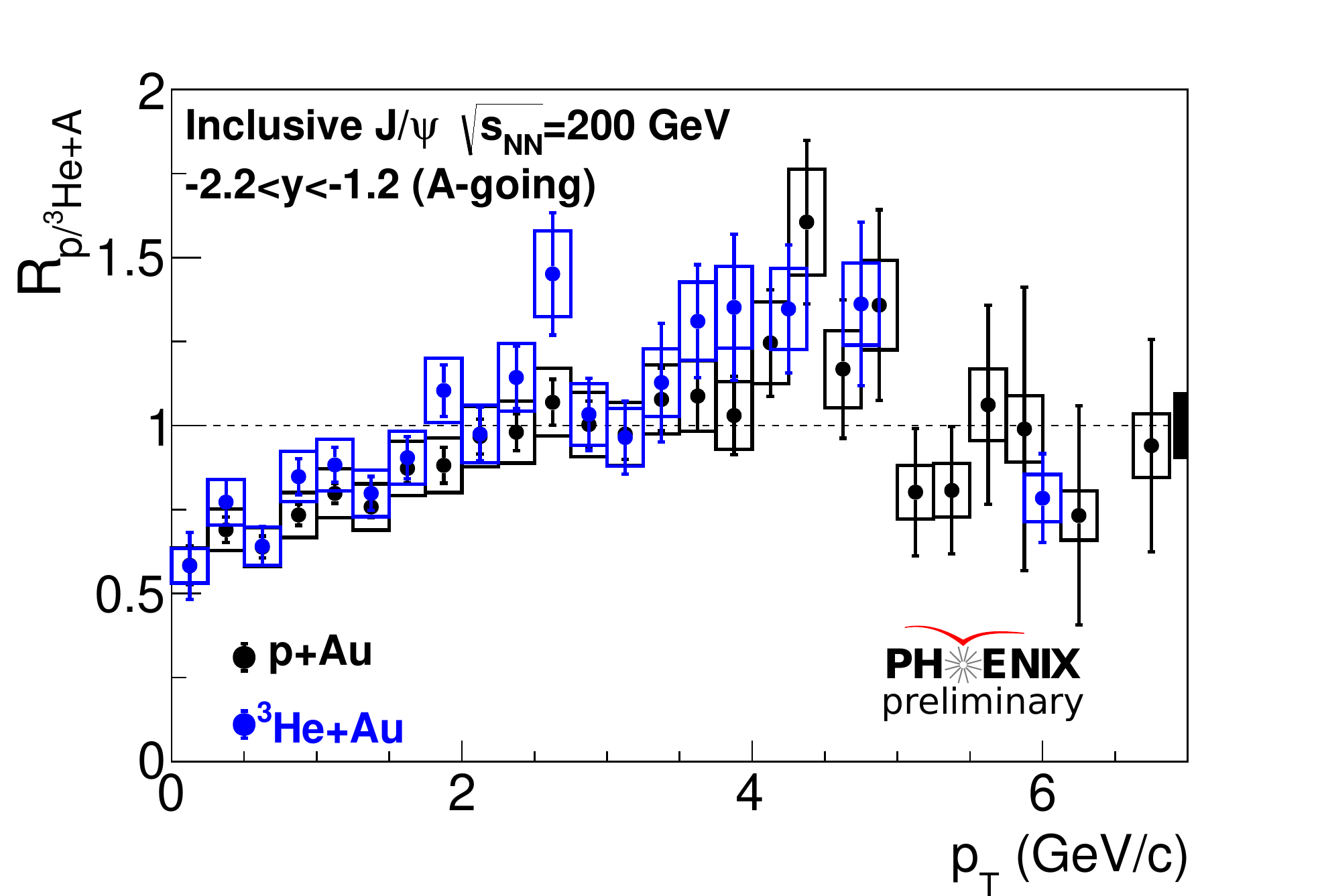}
    \includegraphics[width=0.495\textwidth]{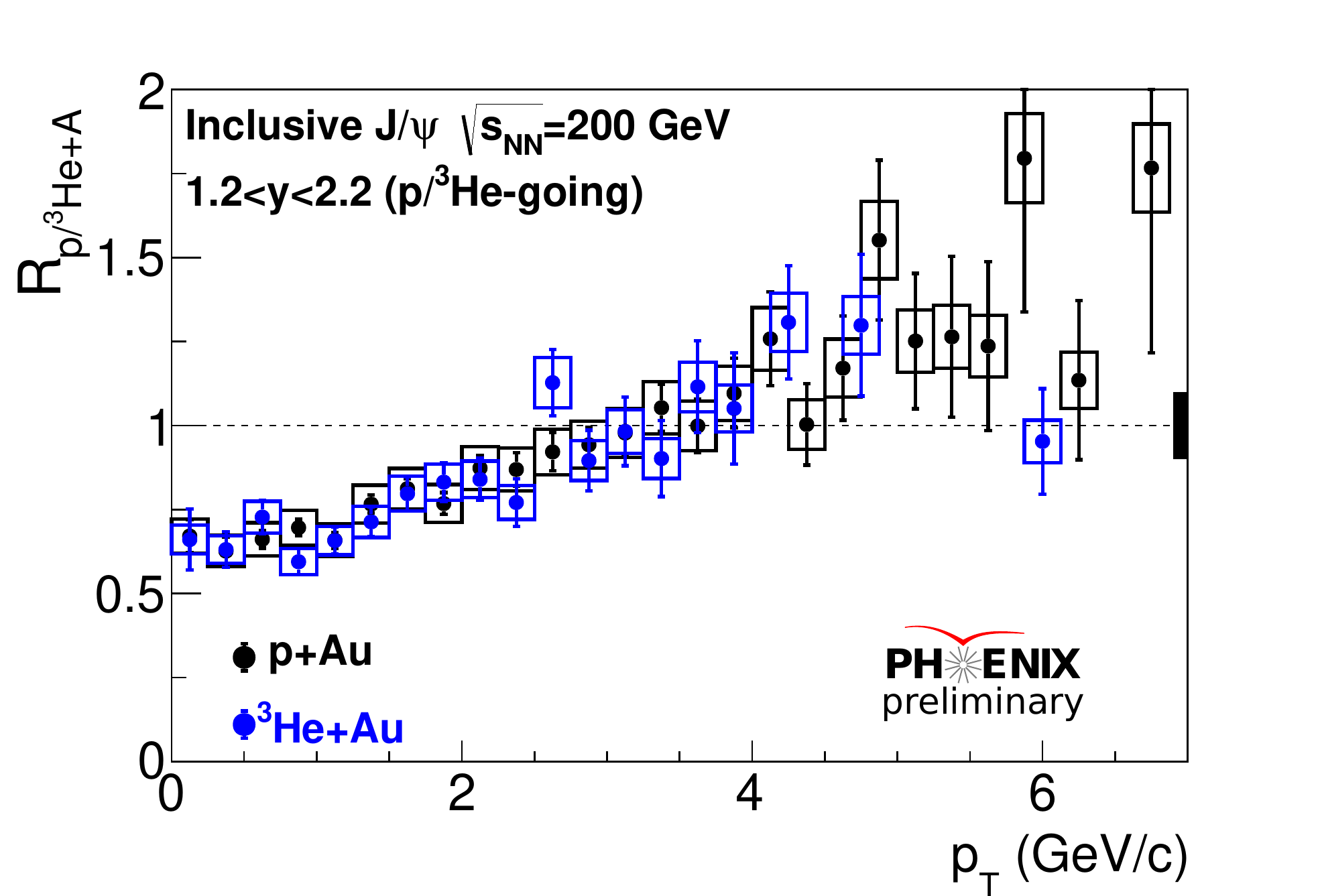}
    \caption{The nuclear modification factor of $J/\psi$ mesons produced at backward (left) and forward (right) rapidity in $p$+Au and $^3$He+Au collisions at 200 GeV.}
    \label{fig:RpAu}
\end{figure}

This is in contrast to the larger $p$+Au and ${^3}$He+Au systems, where significant modification is observed at both forward and backward rapidity.  At backward rapidity, in the Au-going direction (left panel of Fig. \ref{fig:RpAu}), we see a suppression at low $p_{T}$ that disappears with increasing $p_{T}$.  It is interesting to note that the modification is the same (within uncertainties) for both systems, despite the factor of $\sim$2 difference in the number of produced charged particles $dN_{ch}/d\eta$ between these system \cite{PPG221}.  This may suggest that, over this range of $dN_{ch}/d\eta$, late stage breakup of $J/\psi$ is not sensitive to the co-moving particle density within these uncertainties.

At forward rapidity, Fig. \ref{fig:RpAu}, right panel, there is a similar structure observed.  Here $J/\psi$ production samples the shadowing region of the gluon parton distribution function inside the nucleus.  Again we see identical behavior for the $p$+Au and $^3$He+Au systems.  Given that we are varying the projectile but not the nuclear target, this could suggest that initial state effects inside the nucleus such as energy loss or nuclear shadowing are the dominant effects on $J/\psi$ production in these small systems.
	
\begin{figure}[htbp]
    \centering
    \includegraphics[width=0.495\textwidth]{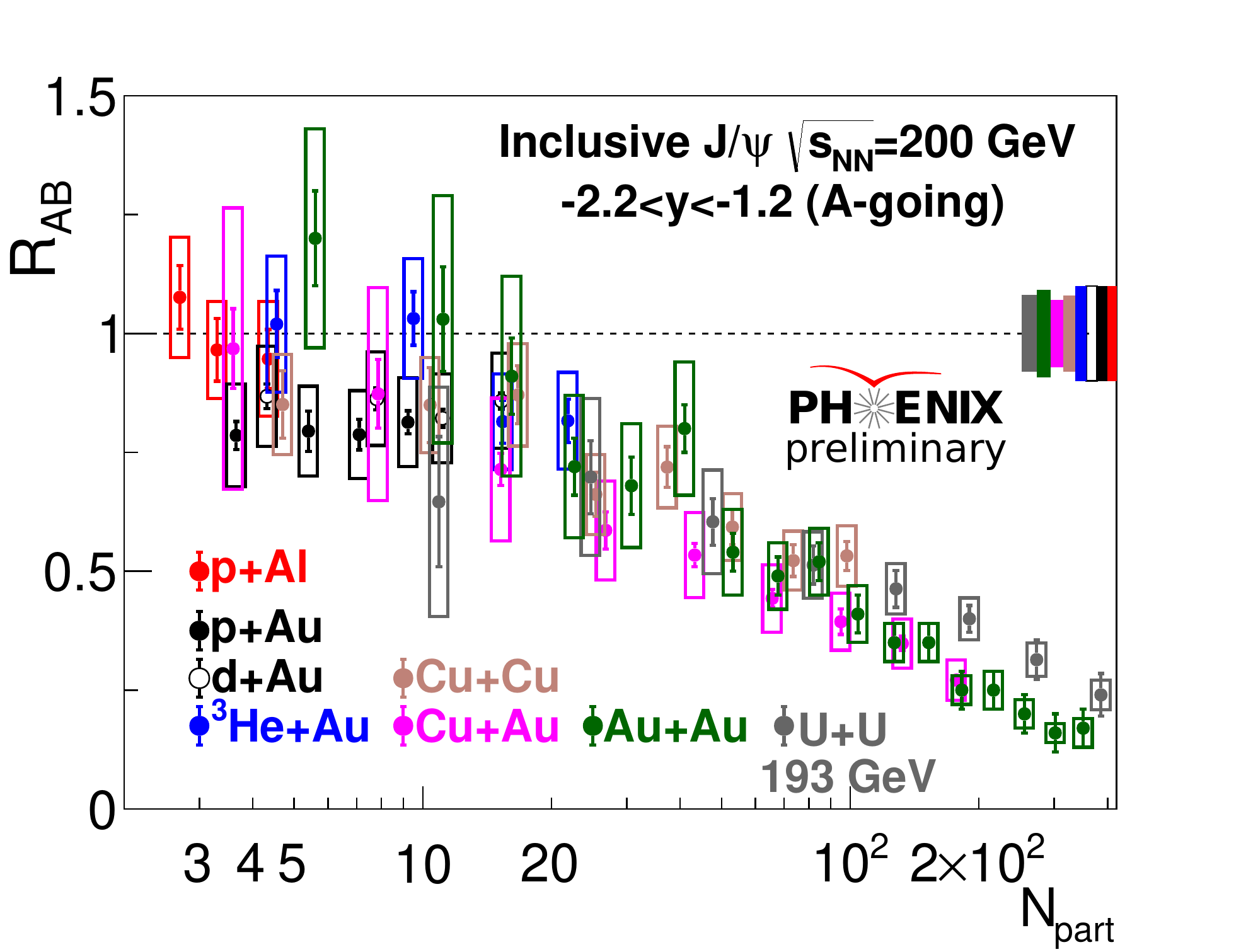}
    \includegraphics[width=0.495\textwidth]{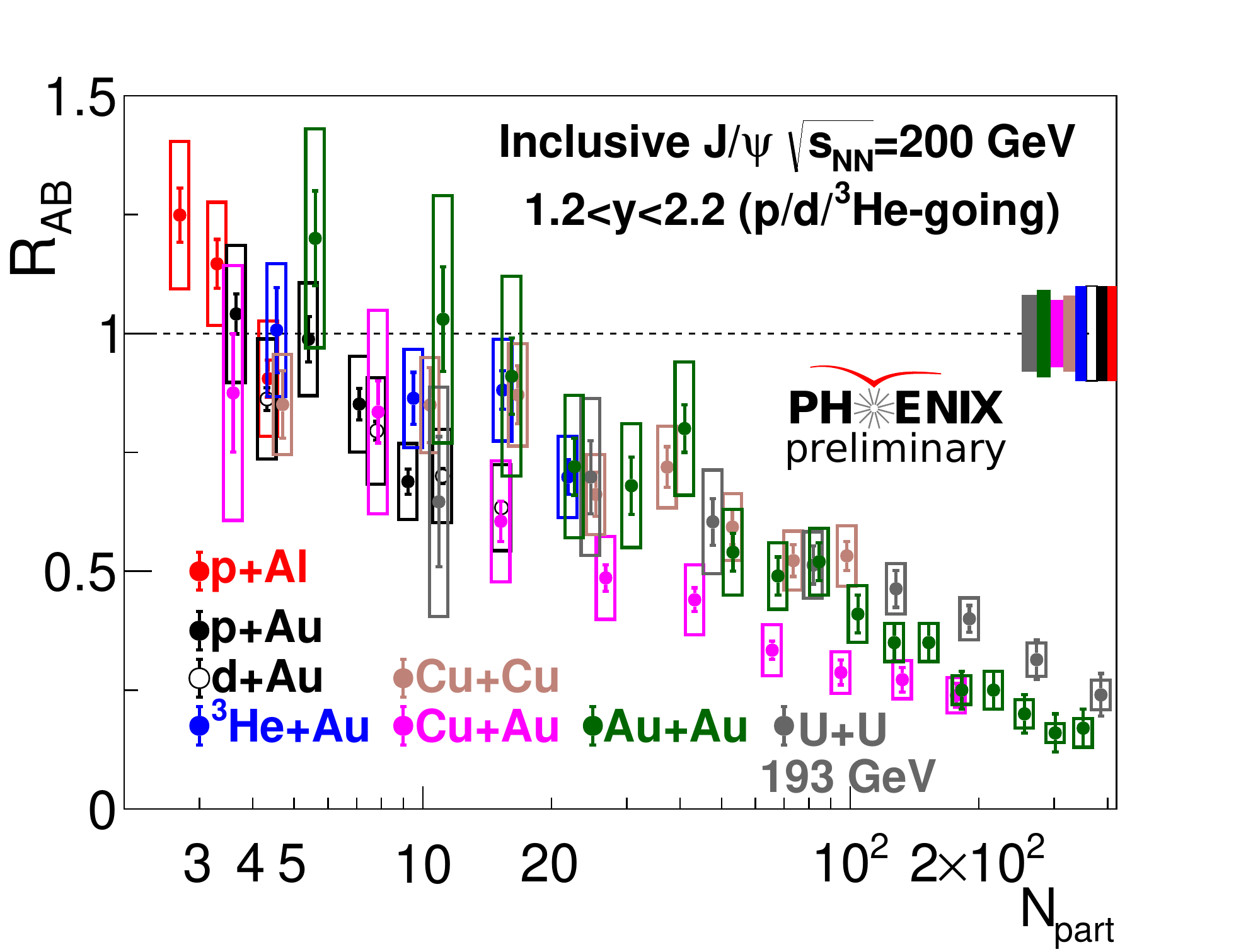}
    \caption{The nuclear modification factor of $J/\psi$ mesons produced at backward (left) and forward (right) rapidity in $p$+Al, $p$+Au, $^3$He+Au, Cu+Cu, Cu+Au, Au+Au, and U+U collisions at 200 GeV.}
    \label{fig:RpAll}
\end{figure}

We now compare these new results in small systems to existing data on larger collision systems, using PHENIX data recorded in Cu+Cu, Cu+Au, Au+Au, and U+U collisions at RHIC \cite{PPG071, PPG163, PPG119, PPG172}.  The data are shown in the two panels of Fig. \ref{fig:RpAll}, where the same data is plotted in each rapidity range for symmetric systems.   The $p_{T}$-integrated results from small systems are compared with these larger systems as a function of $N_{part}$.  Within uncertainties, we observe a similar trend across all systems, with a large suppression developing with increasing system size.  Taken together, these data probe a very large range of system size, temperature, and energy density.

\section{Summary}

	Quarkonia measurements have been a cornerstone of the PHENIX physics program from the earliest days of RHIC.  With these results, we have additional data that can constrain models of charmonium production in small systems, which allows us to quantify effects which are crucial for fully understanding the role color screening may play in the significant $J/\psi$ suppression observed in large systems.  While PHENIX is no longer recording data, analysis of the large Au+Au data sets recorded in 2014 and 2016 is currently underway.

\bibliographystyle{JHEP}
\bibliography{PPG188}


\end{document}